\documentclass[english,prl,twocolumn,showpacs,floatfix]{revtex4}
\usepackage[T1]{fontenc} \usepackage[latin1]{inputenc}
\usepackage{babel} \usepackage{epsfig}
\usepackage{amsfonts}

\begin{document}

\title{A novel approach to synchronization in coupled excitable
systems} \author{B.~Naundorf} \affiliation{Max-Planck-Institut f\"ur
Str\"omungsforschung and Institut f\"ur Nichtlineare Dynamik der
Universit\"at G\"ottingen, Bunsenstr.~10, 37073 G\"ottingen, Germany}
\author{T.~Prager, L.~Schimansky-Geier} \affiliation{Institute of
Physics, Humboldt-University of Berlin, Invalidenstr. 110, 10115
Berlin, Germany } \date{\today}

\begin{abstract}
  We consider networks of coupled stochastic oscillators.  When
  coupled we find strong collective
  oscillations, while each unit remains stochastic.  In the limit \(
  N\to \infty \) we derive a system of integro-delay equations and
  show analytically that the collective oscillations persist in a
  large region in parameter. For a
  regular topology with \emph{few} connections between the
  oscillators, islands of coherent oscillations are formed, which
  merge as the amount of topological disorder increases. We link
  this transition to typical network quantities in the framework of
  small-world networks.
\end{abstract}

\pacs{05.45.Xt, 05.10.Gg, 84.35+i} 
\keywords{stochastic oscillators, synchronization,
delayed differential equation, small world networks} 
\maketitle

Collective behavior or synchronization of non-equilibrium systems is
one of the main topics in current complex systems research.  It can be
observed in a variety of different physical, biological and
sociological frameworks, such as sub-excitable media, in which
noise-induced coherent patterns emerge~\cite{Zhou}, neurons in the
brain, where spike patterns synchronize during expectancy or 
attention~\cite{EngelFriesSinger}, or in an enthusiastic audience 
which applauds
in synchrony after a good performance~\cite{NedaBarabasi}.

Starting from the pioneering work of Winfree~\cite{Winfree} and
Kuramoto~\cite{Kuramoto,Strogatz2} on coupled phase oscillators,
numerous studies have focussed on systems where the dynamics of the
single units is \emph{deterministic}~\cite{PikovskyBook}. Even in systems consisting of 
coupled chaotic units
phase locking may occur~\cite{Rosenblum}.  However, 
many systems are intrinsically
\emph{stochastic} or driven by noise.  Numerical investigations show,
that synchronization can be found there as 
well~\cite{Nikitin,Kurrer,Hempel,Neiman99}, or that it may even be induced
by a suitable chosen noise strength~\cite{PikovskyKurthsCR,
Lindner_CR_SR}. To really understand why, and under which conditions
synchronization appears in coupled stochastic
systems, analytically tractable models are indispensable. 
There, however, little advancement has been achieved so far.

In this Letter we
introduce a system of $N$ coupled discrete stochastic oscillators,
each oscillator serving as a prototype of an excitable system. 
We find large coherent oscillations of the ensemble.
In the limit \( N\rightarrow \infty \) we show \emph{analytically} that this
coherence is due to a Hopf bifurcation in the dynamics of the ensemble
distributions in case of a global coupling. Furthermore, 
we show numerically, that the synchronization is preserved
even if there are only \emph{few} connections between the oscillators.
In contrast to previous investigations~\cite{Percora,Kaneko} it
becomes most pronounced for a large amount of randomness in the topology of
the network, which we study in the framework of small world 
networks~\cite{WattsStrogatz}.
\begin{figure}[h]
\centerline{\epsfig{file=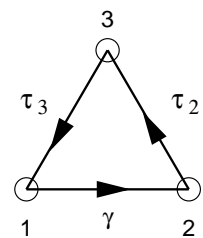,width=3cm,angle=270}}
\caption{Three state model of an excitable unit. The process \(1 \to 2 \) 
is Markovian, while the transitions \( 2\to 3 \) and \( 3\to 1
  \) are deterministic with a fixed waiting time}
\label{Model}
\end{figure}

We consider a system of \( N \) stochastic units. Each unit is a
prototypical model of a stochastic excitable oscillator characterized
by an attracting fix point \( 1 \) from which it escapes by noise to
an excited state \( 2 \). It then performs a long excursion \( 3 \) and
finally returns to its rest state \( 1 \) (Fig.~\ref{Model}). The transition \(
1\rightarrow 2\) is controlled by a rate $\gamma$ which can be
expressed as an Arrhenius-like relation \cite{Lindner_Fhn}.  The
probability to stay the time $t$ 
in $ 1 $ is
thus given by,
\begin{equation}
\label{Formel:PoissonTransition}
w_{1\rightarrow 2} \left( t \right) =\gamma \exp \left( -\gamma
t\right)\,,
\end{equation}
with mean and standard deviation $1/\gamma$. The transitions $ 2\to 3
$ and $ 3\to 1 $ are deterministic with a peaked waiting time
distributions at $ \tau _{2} $ and $\tau _{3}$, respectively:
\begin{equation}
w_{2\rightarrow 3} \left( t \right)=\delta(t-\tau_2)\,,~
w_{3\rightarrow 1} \left(t \right)=\delta(t-\tau_3)\,.
\end{equation}
To demonstrate that the dynamics of each individual unit shows typical features
of excitable systems we inspect its power spectrum.  Setting the
output $s(t)=1$ if the unit is in state \(3\) at time $t$,
and $s(t)=0$, otherwise, the spectrum is then determined employing renewal
theory \cite{Stratonovich}:
\begin{equation}
S(\omega)={\frac4{\omega^2( {1/\gamma} +T)}} {\rm Re} \frac{(1-{{\rm i}\omega
/ \gamma}-{e^{{\rm i}\omega \tau_2}})\left(1-e^{{\rm i}\omega
(T-\tau_2)}\right)}{1-{{\rm i}\omega / \gamma} -e^{{\rm i}\omega
T}}\,.
\label{Formel:spectrum}
\end{equation}
\begin{figure}[h]
  {\centering
  \resizebox*{0.8\columnwidth}{!}{\includegraphics{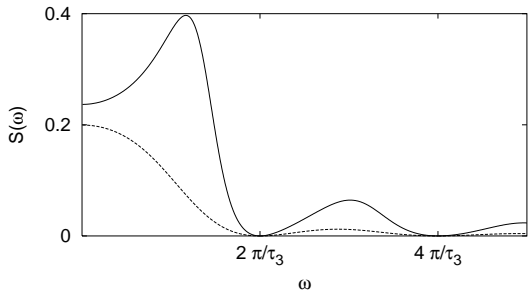}}
    \par} 
\caption{Power spectrum of a single three state unit with $\tau_2=0.1$, $\tau_3=1.2$ and $1/\gamma=1.0$ (solid line), $1/\gamma=10.0$ (dashed line). For small
values of $1/\gamma$ the process is oscillating}
\label{spectrum_unit}
\end{figure}
where $T=\tau_2+\tau_3$.  The stochasticity of the process can be controlled by
varying the amount of time spent in state $1$ compared to $T$.  If
$1/\gamma\gg T$, the system spends a lot of time in state $1$ and
the stochasticity of the first step dominates the dynamics. 
The spectrum decreases then monotonically until it becomes zero at
$\omega_0=2\pi/\tau_3$.  Contrary, for $1/\gamma\ll T$
the process becomes oscillating. (Fig.~\ref{spectrum_unit}).

\begin{figure}[h]
  {\centering
  \resizebox*{0.8\columnwidth}{!}{\includegraphics{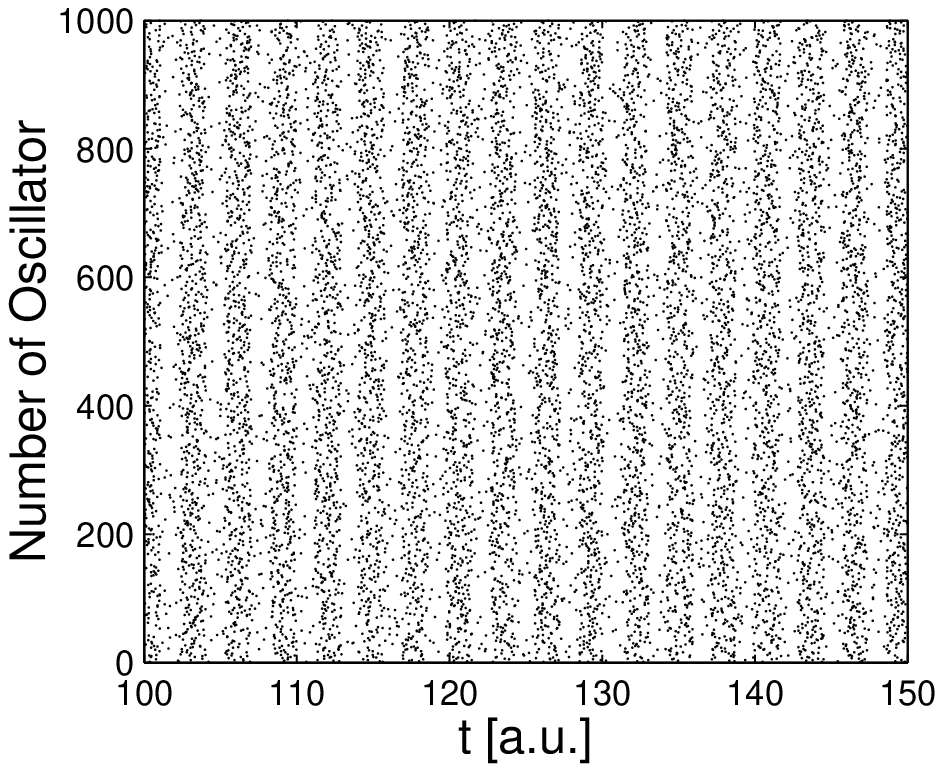}}
    \par} {\centering
    \resizebox*{0.8\columnwidth}{!}{\includegraphics{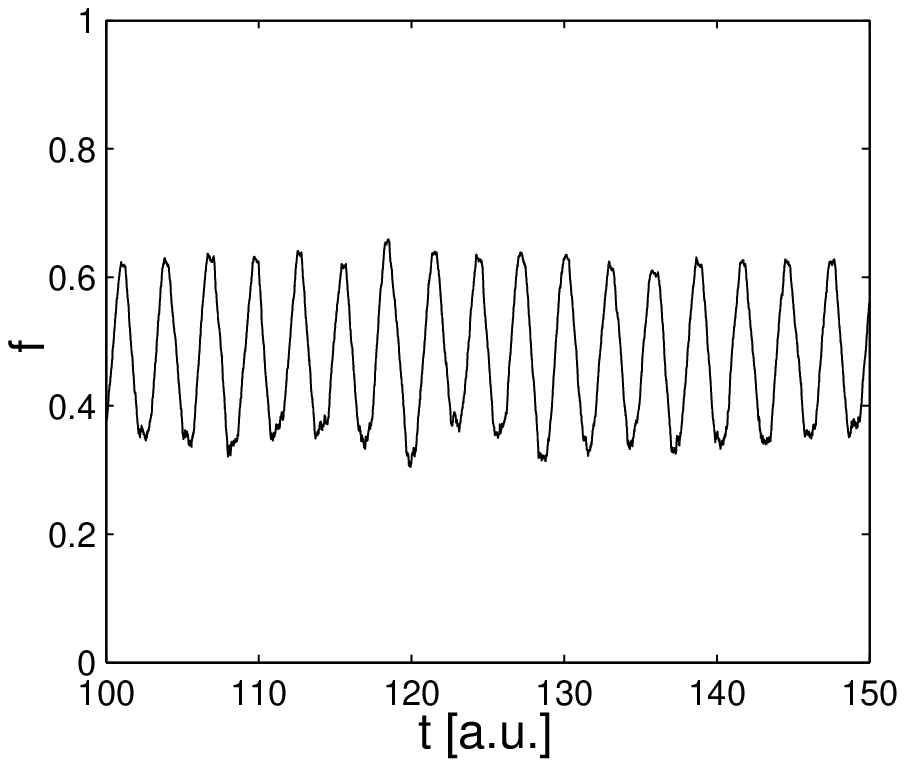}}
    \par}
\caption{\label{Fig:StochOscillations}Spike train for a globally coupled
  network (top) of $ N=1000 $ oscillators. Each time an oscillator
  performs a $ 2\rightarrow 3 $ transition a point is plotted. The
  lower panel shows the dynamic order parameter $ f(t) $.  In all figures the
  following parameters were used: $ T=2.5 $, $ \tau_{2}=0.5 $, $
  \gamma_0=0.5 $, $ \Delta=0.6 $, $ \sigma =10^{-5} $ and $ f^*=0.5 $.
  }
\end{figure}

To couple the units, we first introduce the dynamic order
parameter of the ensemble:
\begin{equation}
f(t)=\frac{1}{N}\sum _{i=1}^{N}s_{i}(t)\, ,
\end{equation}
with $s_i(t)$ denoting the output of unit $i$.  A prototypical
coupling between the individual oscillators is introduced by 
letting the rate $\gamma$ functionally depend on the value of 
$f(t)$ in an inhibitory sigmoidal fashion: 
\begin{equation}
\label{Formel:Rate}
\gamma
(f(t))\,=\,\gamma_0\,\left(1+\Delta\tanh\left[-\frac{f(t)-f^{*}}{
2\sigma} \right]\right)\,.
\end{equation}
Choosing $\Delta >0$, the transition rate is large for small values
and small for large values of $f(t)$.  The position and width of the
transition is mediated by the parameters $ f^{*} $ and $\sigma$. In the
following the position $ f^{*} $ is kept equal to $0.5$. The inverse
of the parameter $\sigma$ can be seen as an effective coupling
parameter. If $\sigma\gg1$, the coupling depends weakly on the
value of $f(t)$ and $\gamma$ is approximately equal to $\gamma_0$.
Contrary, in the case of $\sigma\ll1$, a sharp transition between
the two rates $\gamma_{2/1}=\gamma_0(1\pm \Delta)$ occurs if $f(t)$
crosses $f^*$.  A typical trajectory of $f(t)$ is shown in
Fig.~\ref{Fig:StochOscillations} for a system consisting of $ N=1000 $
units.  Although each individual unit is governed by the stochastic
transition $ 1\rightarrow 2 $, the whole system shows an undamped
oscillation, which resembles coherence
resonance in coupled excitable Fitz-Hugh-Nagumo units
\cite{Hempel,Neiman99}.

The simplicity and discrete nature of our model allows to derive an
analytic condition from an integro-delay equation for the onset of the
coherent oscillations. In the limit $ N\rightarrow \infty $ the state
of the system can be described by the ensemble averaged occupation
probabilities $P_{k}(t)$, i.e. the probability that a unit is in
 state $k$ at time $t$. Following a mean field assumption, we
identify the order parameter $ f(t) $ with $ P_{3}(t) $. 
The dynamics is then described by the
following set of integro-delay equations
\begin{eqnarray}
  P_{1}(t) & = & 1-P_{2}(t)-P_{3}(t)\nonumber \label{Formel:ddeq1} \\
  P_{2}(t) & = & \int _{t-\tau _{2}}^{t}\gamma (P_{3}(t'))P_{1}(t')\,
  dt'\label{Formel:ddeqs} \\ P_{3}(t) & = & \int _{t-T}^{t-\tau
  _{2}}\gamma (P_{3}(t'))P_{1}(t')\, dt'\nonumber\,.
\end{eqnarray}
While the first equation expresses the normalization condition, the
second and third account for the balance of probability. The
probabilities $ P_{2}(t) $ and $ P_{3}(t) $ 
are equal to the time-integrated decay from
state $ 1 $ from time $ t-\tau _{2} $ up to $ t $ and
 from $t-T$ up to $t-\tau_2$, respectively. 
Similar equations have been investigated before in the framework of
delayed differential equations \cite{glasslongtin}. For a two state
rate equation,
however, only a resonant-like behavior has been found \cite{ohira}. 

Now we consider the unique stationary fix-point of the nonlinear delay system
(\ref{Formel:ddeq1}) by setting $ P_{k}(t)=P^{*}_{k} $ \cite{uniqueFP}.  This leads to
an implicit relation for e.g.~$ P_{3}^{*} $
\begin{equation}
\label{Formel:SteadyState}
P_{3}^{*}=\frac{\tau _{3}}{T+1/\gamma \left( P_{3}^{*}\right) }\,,
\end{equation}
which is the ratio between the time spent in state $3$ and the mean
time for one round trip. Analogous relations may be derived for $
P_{1}^{*} $ and $ P_{2}^{*} $.  To investigate the stability of the
single steady state we add small
perturbations $P_{k}(t)=P_{k}^{*}+\delta P_{k}(t)$ with
$\sum_{k=1}^3\delta P_k=0$.  Using $ \delta P_{k}(t) \propto
\exp(\lambda t) $ with $\lambda \ne 0$ leads after linearization of
Eqs.~(\ref{Formel:ddeqs}) 
to the characteristic equation:
\begin{equation}
\label{Formel:CharakteristischeGl}
1+\lambda^{-1} \left\{ \gamma(P_{3}^*)\left(1- e^{-\lambda T}\right)
-s \left( e^{-\lambda \tau_2}-e^{-\lambda T}\right) \right\} =0.
\end{equation}
Here we have introduced $s=\gamma '(P_{3}^{*})P_{1}^{*} <0$ and
$\gamma^{\prime}(\cdot)$ as the first derivative of $\gamma(\cdot)$.

\begin{figure}[h]
  {\centering
  \resizebox*{0.8\columnwidth}{!}{\includegraphics{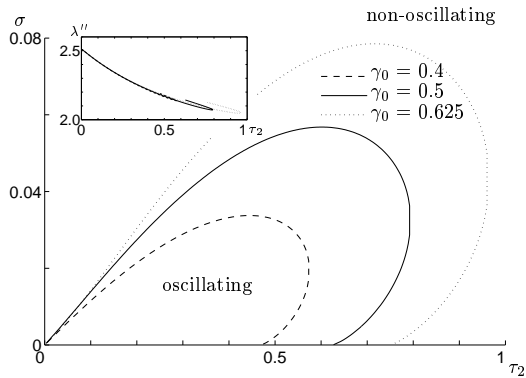}}
    \par}
\caption{\label{Fig:BifurcationDiagram_tau} Bifurcation diagram showing 
  the parameter region of coherent oscillations in the
  $\tau_2$-$\sigma$ plane for different values of $\gamma_0$. The
  inset shows the corresponding frequency at the
  bifurcation, which is nearly independent of $\gamma_0$.  
  All other parameters as in
  Fig.~\ref{Fig:StochOscillations}}
\end{figure}

\begin{figure}[h!]
  {\centering
    \resizebox*{0.8\columnwidth}{!}{\includegraphics{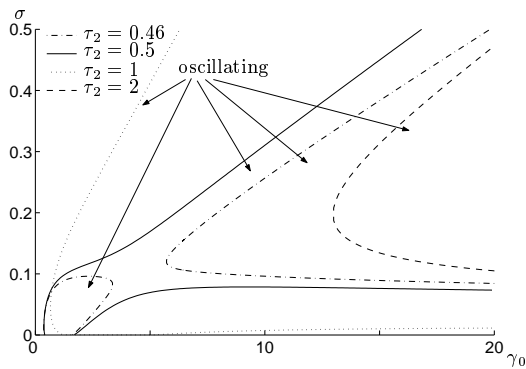}}
    \par}
\caption{\label{Fig:BifurcationDiagram_gamma}Bifurcation diagram showing the parameter region of coherent
  oscillations in the $\gamma_0$-$\sigma$ plane for different values of
  $\tau_2$.
  All other parameters as
  in Fig.~\ref{Fig:StochOscillations}}
\end{figure}

The solutions of Eq.~(\ref{Formel:CharakteristischeGl}) in
$\lambda=\lambda^{\prime}+{\rm i}\lambda^{\prime \prime}$, 
$\lambda',\lambda''\in \mathbb R$ are complex. A Hopf bifurcation
corresponds to values where $\lambda$
crosses the imaginary axis, for which we can derive a 
condition by setting $\lambda^\prime=0$. 
This gives in parametric dependence on the
frequency $\lambda^{\prime\prime}$ at the bifurcation:
\begin{eqnarray}
&&\gamma(P_3^*)\left(1-\cos(\lambda^{\prime \prime}
T)\right)\,-\,s\,\left(\cos(\lambda^{\prime \prime}
\tau_2)-\cos(\lambda^{\prime \prime} T)\right)\,=\,0\nonumber \\ &&
\lambda^{\prime \prime}\,+\,\gamma(P_3^*)\,\sin(\lambda^{\prime
\prime} T)\,+\,s\,\left(\sin(\lambda^{\prime \prime}
\tau_2)-\sin(\lambda^{\prime \prime} T)\right)\,=\,0\,.\nonumber
\end{eqnarray} 
Figs.~\ref{Fig:BifurcationDiagram_tau} and
\ref{Fig:BifurcationDiagram_gamma} shows the region of coherent
oscillations in the $ \tau _{2}-\sigma $ and $\gamma_0-\sigma $ plane,
respectively.
For $\tau_2>0$, fixed $\gamma_0$ and $\sigma\ne 0$ there is a large
region in which the ensemble oscillates. This region grows for
increasing values of $\gamma_0$. Interestingly, there exist values of $\tau_2$
where $\sigma$ has to exceed some finite value to observe the
oscillations. 
When uncoupled, the power spectra of the single units exhibit a maximum
at finite frequency for the parameters given by the three curves shown 
in Fig.~\ref{Fig:BifurcationDiagram_tau}.
However, we would like to stress 
that coherent oscillations of the ensemble can also be observed if the 
single uncoupled units are in the non-oscillating regime.
The frequency at the bifurcation, as shown in the inset of 
Fig.~\ref{Fig:BifurcationDiagram_tau},
which is about $2.2/(2\pi)\approx 0.35$ for $\tau_2=0.5$
agrees well with the
frequency of the oscillations observed in the numerical simulation 
(Fig.~\ref{Fig:StochOscillations}) which is roughly $0.34$.
The dependence on $\gamma_0$ for different values of $\tau_2$ is more 
complex, which we
show in Fig.~\ref{Fig:BifurcationDiagram_gamma}. Surprisingly, there are
two separated regions of coherent oscillations for small
values of $\tau_2$ in the
parameter plane (dash-dotted lines). This means that a path in the 
parameter plane in the direction of increasing 
randomness, i.e.~ 
decreasing  $\gamma_0$, may cause the oscillations first to vanish and 
then reappear.
Upon increase of $\tau_2$ the two regions grow and merge for 
$\tau_2\approx0.48$ (solid line). The merged region maximizes 
at $\tau_2 \approx 1$ 
(dotted line), further increase of $\tau_2$ lets it shrink  
again (dashed line). 
In this case, only large values of $\gamma_0$ 
above $\approx14$
lead to global oscillations.

\begin{figure}[h]
{\centering
\resizebox*{0.9\columnwidth}{!}{\includegraphics{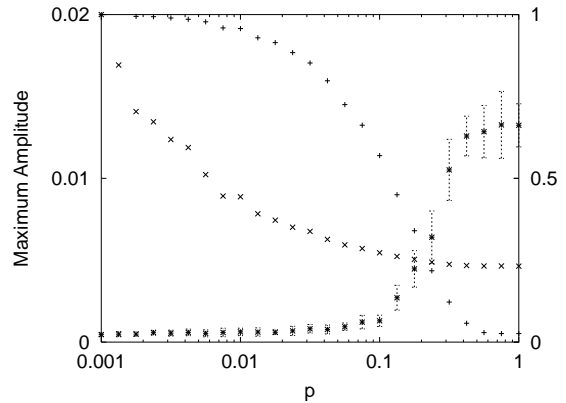}}
\par}
\caption{\label{Figure:SpectrumSW}Left axis: Location of the maximum amplitude of the spectrum of the
  dynamic order parameter  $ f\protect $ ($+\hskip-7.2pt\times$). Right axis: Cluster index (+)
  and mean shortest path length ($ \times \protect $) in the
  network.}
\end{figure}

To complement the globally coupled case, we now
investigate diluted networks and the influence of the topology on the
onset of coherent oscillations in our system. To study this, we
consider a transition from an ordered topology to a random network via
small-world networks, which have recently been introduced by Watts
and Strogatz \cite{WattsStrogatz}.
The transition is constructed as follows: Starting from a ring with $
N $ vertices, each vertex identified as one three state unit, is
coupled to its $ k $ nearest neighbors with undirected edges. This 
means, that the 
transition rate $\gamma_j(\cdot)$ of unit $j$ now depends on the local 
mean field:
\begin{equation}
f_j(t)=\frac{1}{k} \sum_{\{i,j\}} s_i(t),
\end{equation}
$\{i,j\}$ denoting the set of neighbors of vertex $j$.
With
probability $ p $ each edge is then cut and reconnected to a randomly
chosen different vertex. In this way the parameter $ p $ interpolates
between a completely regular ($p=0$) and a random network
($p=1$). Please note that the number of connections is $ Nk $ which we
require now to be much smaller than the number of all possible
connections between vertices, which is $ N(N-1)/2 $. The small world
networks are found for values of $p$ below $\approx0.1$, where the 
mean shortest
path between two arbitrary nodes drops rapidly, while the cluster
index, giving the relative number of common neighbors, is still large (see
Fig.~\ref{Figure:SpectrumSW}).

\begin{figure}[h]
{\centering
\resizebox*{0.8\columnwidth}{!}{\includegraphics{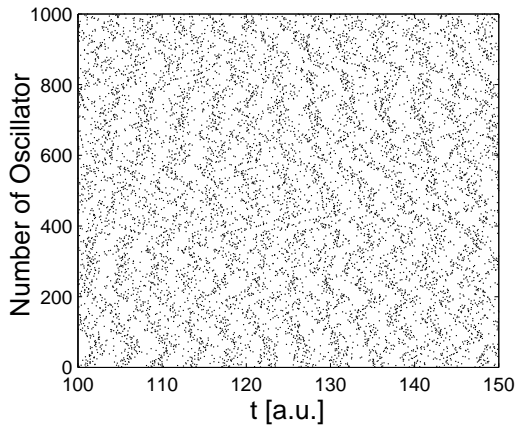}}
\par}
\caption{\label{Fig:SpikeTrainp=3D0}Spike train for a regular network with $p=0$, $N=1000$ and $k=18$. Each time a stochastic unit undergoes a $2\rightarrow 3$ transition a point is plotted. Other parameters are the same as in Fig.~\ref{Fig:StochOscillations}}.
\end{figure}

In Fig.~\ref{Fig:SpikeTrainp=3D0} we show the dynamics on a completely
regular network, i.e.~$p=0$.
There, we find small coherent islands, which interchange over time.
Since these islands are not in phase, there are no \emph{global}
oscillations.
As a measure for the size of the global oscillations, we choose the
height of the central peak in the power spectrum of $f(t)$. For
increasing randomness, i.e.~$ p>0 $, keeping $ k $ fixed, the
oscillations become more pronounced and are finally maximized for
complete disorder, which can be seen in Fig.~\ref{Figure:SpectrumSW}.
The transition to macroscopic oscillations shows a threshold behavior
at $p \approx 0.2$. In the inset, both the cluster index and the mean
path length are shown.  Although there is a steep decrease in the mean
path length for already small $ p \approx 0.01 $, the amplitude of the
oscillations only starts to increase, as the cluster index becomes
smaller. This means, that in our model, completely random networks
synchronize best.  For the considered stochastic systems this stands
in clear contrast to a previous study of chaotic systems \cite{Percora},
where it was claimed that the small-world route provides better
synchronizability, compared to completely random graphs.

In conclusion we have presented a system of $ N $ discrete stochastic
units, which constitutes a generic model of coupled excitable
systems. When coupled we observed coherent
oscillations. In the limit $N\rightarrow \infty $ the system can be
described by a system of integro-delay equations.  A stability
analysis reveals, that the fixed point of the dynamics becomes
unstable under variation of the system parameters. We argue that this
mechanism is valid for the finite system as presented by
simulations.

We further showed numerically, that the coherent oscillations are
preserved if the connectivity is strongly diluted. Starting from a
regular network with $ k $ nearest neighbor connections ($ kN\ll
(N-1)N/2 $) islands of coherent oscillations with different relative
phases coexist.  With topological disorder these islands merge and
macroscopic oscillations of the whole ensemble can be observed. The
transition to global oscillations is connected to the cluster index in
the network, which drops considerably on the onset of the
oscillations.

The authors thank F.~Wolf, M.~Timme, M.~Zaks and J.~Freund for useful
discussion. This work has been supported by DFG Sfb-555.


\begin{thebibliography}{10}
\bibitem{uniqueFP} The uniqueness of the fix point is due to the chosen
  coupling function in Eq.~(\ref{Formel:Rate}), which leads to only one solution
  of Eq.~(\ref{Formel:SteadyState}).
\end{thebibliography}
\end{document}